# Educators on the Frontline: Philosophical and Realistic Perspectives on Integrating ChatGPT into the Learning Space

Surajit Das[1], Peu Majumder[2], Aleksei Eliseev[3]

ITMO University, Russia (mr.surajitdas@gmail.com)[1]
University of North Bengal, India (peumajumder@gmail.com)[2]
Moscow Institute of Physics and Technology, Russia (eliseev.av@mipt.ru)[3]

**Abstract:**

The rapid emergence of Generative AI, particularly ChatGPT, has sparked a global debate on the future of education, often characterized by alarmism and speculation. Moving beyond this, this study investigates the structured, grounded perspectives of a key stakeholder group: university educators. It proposes a novel theoretical model that conceptualizes the educational environment as a "Learning Space" composed of seven subspaces to systematically identify the impact of AI integration. This framework was operationalized through a quantitative survey of 140 Russian university educators, with responses analyzed using a binary flagging system to measure acceptance across key indicators. The results reveal a strong but conditional consensus: a majority of educators support ChatGPT's integration, contingent upon crucial factors such as the transformation of assessment methods and the availability of plagiarism detection tools. However, significant concerns persist regarding its impact on critical thinking. Educators largely reject the notion that AI diminishes their importance, viewing their role as evolving from information-deliverer to facilitator of critical engagement. The study concludes that ChatGPT acts less as a destroyer of education and more as a catalyst for its necessary evolution, and proposes the PIPE Model (Pedagogy, Infrastructure, Policy, Education) as a strategic framework for its responsible integration. This research provides a data-driven, model-based analysis of educator attitudes, offering a nuanced alternative to the polarized discourse surrounding AI in education.

Introduction

The advent of Generative Artificial Intelligence (GenAI), particularly with the release of OpenAI's ChatGPT in late 2022, has ignited a global discourse on the future of various human domains, with education standing at the epicenter of this transformation. Capable of generating sophisticated text, code, and analyses, these large language models present a paradigm shift, challenging centuries-old pedagogical traditions and institutional structures. The rapid integration of such tools into the academic sphere has provoked a spectrum of reactions—from enthusiastic adoption to profound apprehension.

A central point of contention lies in the perceived threat to the core mission of education itself. Critics voice concerns that GenAI facilitates academic dishonesty, promotes superficial learning, and potentially undermines the development of critical thinking skills essential for a thriving society. This perspective often frames AI as a disruptive force that could devalue the role of educators and reduce learning to a transaction with an algorithm. However, this alarmist view often rests on a narrow definition of education, one conflated with rote memorization and high-stakes standardized testing—a system already facing scrutiny for its limitations in fostering genuine understanding and human values.

This debate necessitates a return to the first principles: What is the fundamental purpose of education? As recognized by the United Nations as a fundamental human right, education is more than accumulation of facts; it is a holistic process of acquiring knowledge, skills, values, and beliefs that empower individuals and contribute to societal development. From this philosophical standpoint, the true crisis in education may not be heralded by AI, but rather exposed by it. If an AI tool can easily circumvent an assessment, the flaw may lie not with the tool, but with the assessment's design, which fails to measure deeper cognitive and ethical capacities that define an educated mind.

The response from one of the primary stakeholders in this transition—educators—is crucial, yet systematically underexplored, especially within specific cultural contexts like Russia. While a growing body of literature catalogues the functional opportunities and threats of ChatGPT, few studies ground this analysis in a structured theoretical model of the learning environment or quantitatively capture the attitudes of those tasked with implementing this change in the classroom.

This study, therefore, seeks to investigate the philosophical and realistic perspectives of Russian university educators on the integration of Generative AI into the learning space. However, to move beyond a simplistic cataloguing of opinions, this research is grounded on a novel theoretical construct: **the "Learning Space" model.**

The development of this model is essential and critical for three primary reasons. First, the discourse on AI in education often remains fragmented, analyzing impacts on assessment, pedagogy, or ethics in isolation. Our model provides a **comprehensive and systematic framework** that conceptualizes the educational environment as an ecosystem of seven interconnected subspaces—from Learning Theory to Infrastructure and Assessment. This allows for a holistic diagnosis of AI's impact across the entire educational system, rather than on isolated parts.

Second, the model **bridges a critical gap between philosophical debate and empirical measurement.** It operationalizes abstract educational values and complex pedagogical relationships into concrete **identifiers** and measurable **indicators**. This enables us to move from asking "Is AI good or bad?" to quantitatively investigating "How does AI affect specific, defined components of valuable education?"

Finally, by anchoring this model in established theories like Social Learning Theory and Actor-Network Theory, we ensure that our investigation contributes to broader scholarly conversations, providing a structured lens through which the transformative role of AI can be systematically analyzed and understood across different contexts.

Guided by this framework, the research is driven by the following questions:

1. Do educators perceive Generative AI as a positive or negative force in higher education?
2. What are the key identifiers and indicators within the learning space that can measure the impact of this integration?
3. Is ChatGPT likely to transform education, and if so, in what manner?

By addressing these questions, this research aims to contribute a nuanced, data-driven perspective to the ongoing debate, arguing that the future of education in the age of AI will be shaped not by the technology itself, but by our collective ability to harness it in service of education's deepest and most humanistic goals.

The paper is structured as follows: review of the relevant literature, presentation of the theoretical foundations of the Learning Space model, detailed methodology, presentation of the results, comprehensive discussion of the findings, and concluding section that states the implications and limitations of the study.

**Literature Review**

The rapid proliferation of GenAI, particularly ChatGPT, has triggered a significant shift in educational landscapes worldwide. This technological disruption has compelled educators to re-evaluate long-standing pedagogical practices, assessment methods, and their own professional roles. A growing body of literature has begun to explore these changes, primarily focusing on the perceptions, challenges, and opportunities as seen by key stakeholders, especially educators. This review synthesizes existing research into four central themes: the spectrum of global educator attitudes, the conceptual framing of AI's paradoxical role, the acute focus on assessment reform, and the identification of persistent geographical and methodological gaps. Through this synthesis, the necessity for a model-based, quantitative study within the Russian academic context becomes clear.

A dominant theme in the literature is the exploration of educator perceptions, which reveal a spectrum of attitudes ranging from cautious optimism to profound concern. Early studies, often conducted via surveys and interviews, capture this initial phase of adaptation. For instance, a study with school teachers and leaders in an Eastern cultural context identified impacts across four key domains: learning, teaching, assessment, and administration [Lau & Guo, 2023]. Similarly, a Community of Practice in Australia found educators struggling with outdated assessment methods, exhibiting a wide range of responses from reactionary bans to thoughtful engagement with the technology (Luckin & Holmes, 2024). This sentiment is echoed in a survey of educators recruited via social media, which captured preliminary perceptions but was limited by its self-selecting sample [Yan, 2023]. A smaller-scale study in northern Sweden further confirmed this varied landscape among upper secondary school staff [Westin & Nordin, 2024]. Collectively, these studies establish that the initial educator response is not monolithic but is universally characterized by a state of flux, with a recurring emphasis on the need for institutional guidance and professional development.

Beyond empirical surveys, conceptual research has provided frameworks for understanding the complex tensions introduced by GenAI. These studies posit that the technology's impact is not a simple binary but a series of inherent paradoxes that educators must navigate. For example, Zhang and Wang (2023) framed these tensions as four-fold: Generative AI is a "friend" yet a "foe," "capable" yet "dependent," "accessible" yet "restrictive," and becomes more "popular" when "banned." Another study, grounded on the constructivist theory of learning, analyzed the alignment of ChatGPT with principles of knowledge construction, highlighting the tension between AI-generated content and student-driven learning [Johnson & Liu, 2023]. These theoretical works provide a valuable lens for interpreting empirical data, suggesting that the conflicts reported by educators are not merely practical but are inherent to the technology's dualistic nature.

Within these broad perceptions and conceptual frames, a predominant and immediate concern for educators globally is the profound impact of GenAI on assessment and academic integrity. This has forced a critical re-evaluation of traditional evaluation methods. A quantitative survey of 389 students and 36 educators explicitly compared their attitudes across various assessment scenarios, highlighting the challenges in maintaining integrity [Chen & Lee, 2023]. The urgency of this issue is particularly acute in teacher training; a qualitative study with language teacher educators in Hong Kong emphasized the pressing need to redesign assessment tasks to focus on the learning process and contextual experiences to mitigate misuse [Wong & Chan, 2024]. This focus on transforming assessment to foster higher-order skills was also a key recommendation in a rapid review of the early literature, though it noted that such suggestions were often based on intuitive beliefs rather than empirical evidence [Holmes & Luckin, 2023]. This theme underscores that assessment is the primary frontline where the tensions of Generative AI are being played out, with a growing consensus that systemic reform is inevitable.

Despite this expanding body of research, significant geographical and methodological gaps persist. The literature remains heavily weighted towards specific contexts, with multiple reviews noting a bias towards Western perspectives [Holmes & Luckin, 2023]. While recent studies have begun to address this in regions like the Arab world [Sadriwala & Al-Salmi, 2025; Al-Saadi, 2025], the perspective of educators in other major academic regions, such as Russia, remains notably under-represented. Methodologically, there is a reliance on qualitative case studies and conceptual analyses, with many studies explicitly acknowledging their lack of empirical data or generalizability [Zhang & Wang, 2023; Selwyn & Jandrić, 2023]. A rapid review of the literature further confirmed that early studies were largely based on preprints and non-empirical suggestions [Holmes & Luckin, 2023]. This points to a critical gap: the lack of a structured, model-based approach to quantitatively measure educator acceptance across defined dimensions of the learning ecosystem.

In conclusion, while the existing literature provides valuable insights into complex and often paradoxical perceptions of educators towards ChatGPT, it is marked by a contextual skew and a predominance of conceptual or small-scale qualitative methodologies. The absence of a unified theoretical model to quantitatively assess the impact of GenAI across different subspaces of the learning environment presents a clear opportunity for further research. This study aims to address this gap by introducing a model of the learning space and presenting the results of a quantitative survey conducted among Russian university educators to systematically measure their acceptance of ChatGPT across key pedagogical identifiers and indicators.

## 2. Theoretical Foundations: A Model of the Learning Space

To move beyond a superficial analysis of ChatGPT as a mere tool, this study is grounded on a novel theoretical construct: the *Learning Space*. This construct serves as an analytical framework to deconstruct the complex ecosystem of education into interdependent subspaces, allowing for a systematic investigation of how Generative AI disrupts, reconfigures, and integrates into existing pedagogical structures. The model posits that learning is not a simple transaction but an emergent property of a complex system, and that the integration of a powerful new actant like ChatGPT necessitates an examination of its impact across this entire system.

### 2.1 Conceptual Rationale and Model Architecture

The Learning Space is defined as the multidimensional environment where learning processes, interactions, and outcomes coalesce. It encompasses not only physical and digital infrastructure but also dynamic interplay of pedagogical theories, stakeholder relationships, and evaluative mechanisms. Our model divides this space into seven non-intersecting subspaces, not as rigid silos, but as interconnected analytical categories. This choice is inspired by systems theory in education, which emphasizes that changes in one part of the system inevitably ripple through others. The introduction of Generative AI represents a systemic perturbation, and this model provides the framework to trace its effects. We use identifiers and indicators within each subspace function as the sensors to measure these effects, and consider many-to-many correspondence between them reflecting non-linear, interconnected nature of educational phenomena.

## 2.2 Elaboration of Subspaces and Their Theoretical Underpinnings

### 2.2.1 Learning Theory and Andragogy: The Case for Social Learning Theory (SLT)

The subspace of Learning Theory and Andragogy provides the pedagogical foundation for the model. While numerous learning theories exist, this study strategically selects **Social Learning Theory (SLT)** as a primary identifier for its unique relevance in the age of AI. Developed by Albert Bandura, SLT posits that people learn not only through direct experience but also through observing the behavior of others and its consequences [Bandura, 1977]. Traditionally, the "model" in SLT is a human peer, teacher, or media figure.

We propose a critical extension. In our interpretation advanced AI chatbots like ChatGPT can function as a **personified, interactive model**. When a student engages with ChatGPT, they are not merely retrieving information; they are observing its language patterns, its problem-solving steps, and its structuring of knowledge. This process of *observational learning* is mediated by cognitive processes—the learner assesses the AI's output, decides what to adopt, and integrates it into their own understanding. This makes SLT a powerful lens for analyzing AI integration, as it bridges behaviorist outcomes (the generated text) with the cognitivist processes (critical evaluation, integration) that the interaction provokes. This stands in contrast to purely behaviorist models (focused on stimulus-response) or constructivist models (focused on individual knowledge construction), by capturing the unique "social" interaction between human and machine.

### 2.2.2 Complex Dynamic System of Engagement Variables: The Matthew Effect in AI Integration

This subspace moves beyond static factors to capture the fluid, interconnected nature of learner engagement. We conceptualize engagement not as a set of independent variables, but as a **Complex Dynamic System (CDS)** comprising affective, behavioral, cognitive, and academic dimensions, all influenced by contextual data like demographics and personality.

The proposed relational model (Fig. 1) illustrates a critical phenomenon: all the loops between these variables are reinforcement loops. This signifies a potential **Matthew Effect** [Merton, 1968]—often summarized as "the rich get richer." In this context, students who initially possess slightly better digital literacy, motivation, or access to technology may leverage ChatGPT more effectively. This initial advantage boosts their cognitive and academic engagement (e.g., through better grades or deeper understanding), which in turn reinforces their positive effect and motivates further behavioral interaction with the AI. Conversely, those with initial disadvantages may struggle to use the tool effectively, leading to frustration (negative affect), disengagement, and a widening gap. This CDS framework allows us to hypothesize that ChatGPT will not have a uniform impact but will likely amplify pre-existing inequalities among students if not actively mitigated, a dimension often overlooked in more simplistic analyses [Malik and Hussain, 2023].

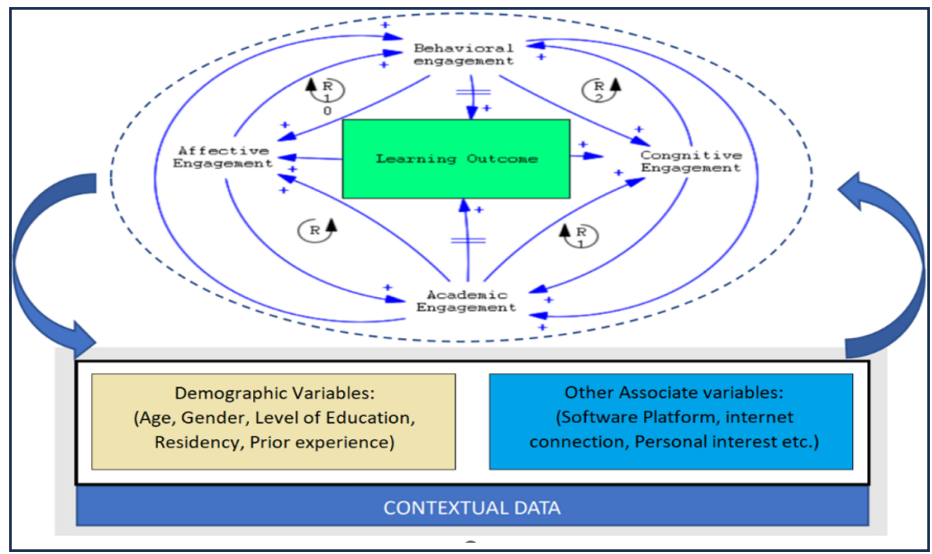

Fig. 1 Causal Diagram of Complex Dynamics System of Engagement Variables

**2.2.3 Stakeholders: Reconfiguring the Network with Non-Human Actants**

The Stakeholders subspace maps the relational ecology of the learning environment. Our graphical representation (Fig. 2) can be powerfully interpreted through the lens of **Actor-Network Theory (ANT)** [Latour, 2005]. ANT argues that agency is not a solely human attribute but is distributed across a network of human and non-human "actants."

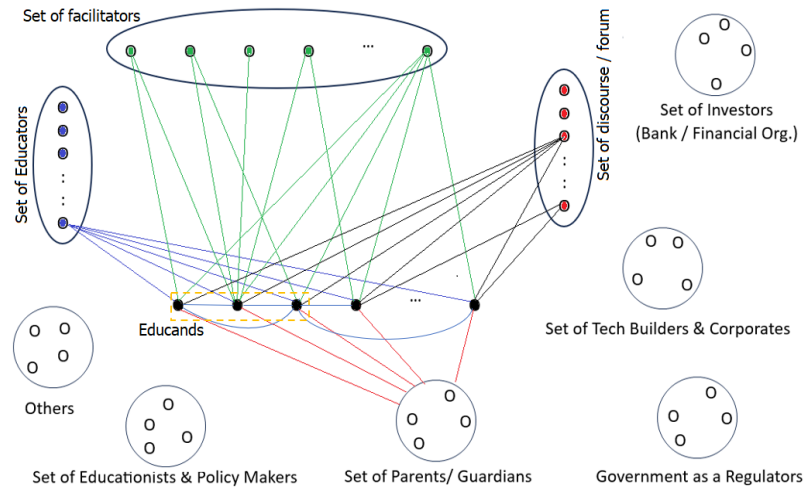

Fig. 2 Relational Diagram of Different Stakeholders

In the traditional network, the primary actants are educators, students, peers, and textbooks. The integration of ChatGPT introduces new non-human actant into this network, which is not a passive tool but an active agent that *mediates* and *transforms* relationships. It changes how students relate to educators (e.g., seeking AI's help before teacher's help), how they relate to peers (changing collaborative dynamics), and how they relate to knowledge itself. The weight of the relation between a learner and ChatGPT, which we measure as an indicator, is a measure of this new actor's influence within the network. This perspective forces a move beyond asking "Do teachers like ChatGPT?" to a more profound question: "How is ChatGPT reconfiguring the entire social fabric of the classroom?"

### 2.2.4 Assessment and Evaluation: Critiquing the "Banking Model" of Education

This subspace is theorized through a critique of what Paulo Freire (1970) termed the "banking model" of education, where students are treated as empty vessels to be filled with knowledge by teachers, and assessments serve as the ledger for these deposits. Traditional, closed-book exam that dominates this model is highly vulnerable to ChatGPT, which can effortlessly "withdraw" and "redeposit" information.

Our indicator advocating for the "Transformation of Assessment" directly challenges this model. Support for open-book, oral, and application-oriented exams signals a shift towards a **constructivist and authentic assessment paradigm**. In this view, the goal is not to audit stored information but to evaluate students' ability to find, critisize, synthesize, and apply knowledge— skills that are increasingly vital and which ChatGPT can actually help to develop, rather than replace. The historical example of Minkowski and Einstein (Question #25—Q#25) serves as a perfect indictment of the banking model's failure to recognize true intellectual potential, reinforcing the need for this transformation.

**2.2.5 The Dark Sub-Space and Null Sub-Space: A Foucaultian and Sociological Lens**

Finally, two conceptual subspaces address the societal and ethical dimensions of AI integration.

The **Dark Sub-Space**, encompassing academic dishonesty, can be analyzed through Michel Foucault's concepts of power and surveillance. The traditional classroom operates on a panoptic model where the teacher's gaze disciplines students into honesty. ChatGPT disrupts this, necessitating a shift to a new, digital panopticon powered by plagiarism detection software. The ethical debate we posed to educators (Q#18) touches directly on this shift in disciplinary power.

The **Null Sub-Space for Myth** addresses the societal-level "moral panic" often accompanying new technologies. The claim that "ChatGPT will destroy critical thinking" is a modern incarnation of this panic, previously aimed at calculators, computers, the internet, and even writing itself (Plato's Phaedrus). By framing this as a subspace, we formally relegate these alarmist, ahistorical claims to the realm of myth, allowing for a more data-driven and calm analysis of the technology's real impacts and challenges.

By anchoring the Learning Space model in these robust theoretical traditions, this study moves beyond a simple report of survey results. It provides a sophisticated, multi-dimensional framework for understanding not just *if* educators accept ChatGPT, but *how* its integration is fundamentally reshaping the very foundations of the educational landscape.

The formal definitions of Subspace, Identifier, and Indicator are provided in the supplementary materials. These materials include Table S-1 (List of Identifiers by Subspace), Table S-2 (Mapping of Identifiers to Their Measured Indicators), and Table S-3 (Educator Survey Questions).

**3. Methodology**

This study employed a quantitative survey design, grounded in the novel theoretical model of the learning space discussed in the previous section. The research was conducted in two integrated phases: (1) the operationalization of the theoretical framework into a structured survey instrument, and (2) the administration of this survey and a subsequent multi-faceted analysis of the collected data to measure educator acceptance against the model's indicators.

**3.1 Survey Instrument Development and Validation**

The survey was designed to translate the abstract identifiers and indicators of the Learning Space model into measurable, empirical data. The development process followed a structured approach to ensure content validity and reliability.

- **Item Generation:** An initial pool of questions was generated, with each question explicitly mapped to a specific indicator within the theoretical model (as summarized in Table 1). The questions were formulated to be clear, concise, and unambiguous, using a close-ended format with predefined response options (typically Yes/No or Agree/Disagree) to facilitate quantitative analysis.

- **Expert Review:** To ensure content validity—that the questions accurately reflected the constructs of the theoretical model—the draft survey was subjected to a review by a panel of three experts in the fields of educational technology, pedagogy, and psychometrics. The panel evaluated the relevance, clarity, and comprehensiveness of each question in relation to its assigned indicator. Their feedback was used to refine question phrasing and eliminate potential ambiguities.

- **Pilot Testing:** The revised survey was then pilot-tested with a small group of 15 educators who were not part of the final sample. The pilot test served to check for average completion time, technical functionality on the Yandex Forms platform, and the clarity of instructions. No major issues were reported, and the instrument was deemed fit for full-scale deployment.

The final survey was structured in Russian and consisted of two main parts: (1) demographic questions, and (2) the core set of questions targeting the model's indicators.

**Table 1: Mapping of Survey Questions to the Theoretical Model**

| # | Identifier (Subspace) | Indicator(s) | Survey Q# |
|---|---|---|---|
| 1 | SLT (Learning Theory and Andragogy) | Personified ChatGPT, Observational Learning | #22 |
| 2 | Behavioural Engagement (Engagement Variables) | Motivation | #19 |
| 3 | Cognitive Engagement (Engagement Variables) | Backward Knowledge Transfer, Critical Thinking | #20, #21 |
| 4 | Traditional Evaluation Process (Assessment) | Flaws in Evaluation System, Transformation of Assessment | #25, #15 |
| 5 | Existence of Educators (Stakeholders) | Importance of Educators | #17 |

| # | Identifier (Subspace) | Indicator(s) | Survey Q# |
|---|---|---|---|
| 6 | Plagiarism Tools (Infrastructural Facts) | ChatGPT Enhancement Requirement, Plagiarism Need | #23, #16 |
| 7 | Dark Sub-Space | Preventing Discrepancy | #18 |
| 8 | Null Sub-Space for Myth | "Nothing is Lost" Philosophy | #24 |

### 3.2 Participants and Sampling Procedure

The target population for this study was educators from Russian universities who are actively involved in mentoring undergraduate, graduate, and postgraduate students. A purposive sampling technique was employed to ensure we captured the perspectives of experienced educators across a range of academic disciplines.

The final sample consisted of **N = 140** educators. The demographic characteristics of the participants are summarized in Table 2, providing a clear picture of the sample's composition. The sample exhibits a strong representation from Engineering and Natural Sciences, with a wide range of teaching experience, suggesting that the findings are grounded in the views of established academics.

**Table 2: Demographic Profile of Survey Participants (N=140)**

| Characteristic | Category | Frequency | Percentage |
|---|---|---|---|
| **Faculty Domain** | Engineering / Natural Science | 89 | 63.6% |
| | Arts & Humanities | 22 | 15.7% |
| | Economics & Management | 16 | 11.4% |
| | Medicine | 9 | 6.4% |
| | Other | 4 | 2.9% |
| **Years of Teaching Experience** | 1-5 years | 18 | 12.9% |
| | 6-10 years | 31 | 22.1% |

| Characteristic | Category | Frequency | Percentage |
|---|---|---|---|
| | 11-20 years | 57 | 40.7% |
| | More than 20 years | 34 | 24.3% |
| **Level of Students Mentored** | Undergraduate | 140 | 100% |
| | Graduate | 112 | 80.0% |
| | Postgraduate | 97 | 69.3% |

**3.3 Data Collection and Ethical Considerations**

Data collection was carried out over a four-week period using the Yandex Forms online platform. The survey link was distributed through professional academic networks and university mailing lists. Participation was voluntary and anonymous, with informed consent obtained from all participants at the beginning of the survey. The survey was configured such that all questions were obligatory, resulting in a complete dataset with no missing values for the core analysis. The anonymized dataset has been made publicly available to ensure transparency and reproducibility.

**3.4 Data Analysis Strategy**

The analysis proceeded in three distinct stages to ensure both a clear overview and a nuanced understanding of the trends.

1. **Descriptive Analysis and Flagging System:** The primary analysis involved calculating the normalized percentage of "favorable" responses for each of the 10 key indicators. A "favorable" response was predefined as the one that supported the constructive integration or highlighted a positive aspect of ChatGPT in relation to the indicator's definition. To provide a clear, overarching measure of acceptance, an **Indicator Flag** was assigned using a predetermined threshold:

    o **Flag = 1:** More than 50% of responses were favorable.

    o **Flag = -1:** Less than 50% of responses were favorable.

    o **Flag = 0:** Exactly 50% of responses were favorable (this case did not occur). This approach is justified for exploratory research aimed at identifying dominant trends and has precedent in similar attitudinal studies in educational technology.

2. **Sub-Group Analysis:** To check the robustness of the findings and explore potential disciplinary differences, the analysis was repeated by grouping the sample into

"Engineering or Natural Science" (n=89) and "Other" faculties (n=51). The flag for each indicator was only finalized if the >50%/<50% consensus held true in both the overall and group-level analyses.

3. **Qualitative Thematic Analysis of Open-Ended Comments:** While the core survey was quantitative, a final open-ended question invited general comments. These responses were analyzed using inductive thematic analysis. Two researchers independently read the responses to identify recurring themes and patterns, which were then used to provide contextual depth and illustrative quotes to explain the quantitative findings.

## 4. Results

The following section presents the findings from the survey, structured according to the analytical strategy. We first provide a comprehensive overview of the indicator flagging, followed by a detailed breakdown of the responses, enriched with qualitative insights from the educators.

### 4.1 Overview of Indicator Flagging

The analysis revealed a strong, conditional consensus in favor of integrating ChatGPT into the learning space. As summarized in Table 3, eight out of ten indicators received positive flag ('1'), indicating that the majority of educators held a favorable view of ChatGPT's impact on those specific dimensions. Two indicators, however, received negative flag ('-1'), pinpointing critical areas of concern. The sub-group analysis confirmed that this trend was consistent across both "Engineering/Science" and "Other" faculties, with no group-level variation altering final indicator flags.

**Table 3: Summary of Indicator Flags from Survey Analysis**

| # | Identifier | Indicator(s) | Survey Q# | Favorable Response | Indicator Flag |
|---|---|---|---|---|---|
| 1 | SLT | Personified ChatGPT & Observational Learning | #22 | 53% | 1 |
| 2 | Behavioural Engagement | Motivation | #19 | 59% | 1 |
| 3 | Cognitive Engagement | Backward Knowledge Transfer | #20 | 72% | 1 |

| # | Identifier | Indicator(s) | Survey Q# | Favorable Response | Indicator Flag |
|---|---|---|---|---|---|
| | | Critical Thinking | #21 | 41% | -1 |
| 4 | Traditional Evaluation Process | Transformation of Assessment | #15 | 71% | 1 |
| | | Flaws in Evaluation System | #25 | 28%* | 1 |
| 5 | Existence of Educators | Importance of Educators | #17 | 29% | -1 |
| 6 | Infrastructural Facts | Plagiarism Need | #16 | 61% | 1 |
| | | ChatGPT Enhancement | #23 | 73% | 1 |
| 7 | Dark Sub-Space | Preventing Discrepancy | #18 | 53% | 1 |
| 8 | Null Sub-Space for Myth | "Nothing is Lost" Philosophy | #24 | 76% | 1 |

*Note: For Q#25, the "favorable" response for the indicator "Flaws in Evaluation System" was the minority (28%). However, this result was interpreted as evidence of the system's failure to mitigate personal bias, thus the flag was set to '1' to indicate a recognized need for systemic transformation.*

## 4.2 Detailed Thematic Analysis of Responses

### 4.2.1 Embracing Pedagogical and Assessment Transformation

A significant majority of educators (71%) supported the transformation of assessment (Q#15), favoring open-book exams followed by an oral component. This sentiment was echoed in qualitative comments, with one educator stating, *"It forces us to ask questions that Google or ChatGPT cannot answer directly, but require synthesis and application. The oral exam then verifies the depth of understanding."* This aligns with the strong consensus (76%) on the historical principle that "nothing is lost" with technological change (Q#24).

### 4.2.2 Conditional Acceptance and Ethical Pragmatism

Educators showed strong but conditional support for integration. The majority (61%) welcomed ChatGPT if supported by reliable plagiarism detection (Q#16), and 73% called for enhanced AI capabilities (Q#23). A pragmatic ethical stance was evident, with 53% permitting the use of ChatGPT for a struggling student in a final retake exam (Q#18). A comment explained this pragmatism: *"Is it better that he leaves the university with nothing, or that he uses a tool to minimally pass and get his degree? The real world will allow him to use these tools, so our assessment must reflect that."*

### 4.2.3 The Evolving Educator and Cognitive Impacts

A pivotal finding was that 71% of educators rejected the idea that AI would diminish their importance (Q#17), resulting in a negative flag for the "existential crisis" indicator. Qualitatively, educators framed their evolving role as follows: *"My job is no longer to deliver information, but to curate, challenge, and inspire. ChatGPT handles the first part; I handle the much more important second part."*

Regarding cognitive impacts, responses were nuanced. While 72% agreed that ChatGPT could facilitate backward knowledge transfer (Q#20), only 41% believed it supported critical thinking (Q#21). This concern was captured in a representative comment: *"I see students accepting the AI's output as dogma. The danger is not in using it, but in stopping the thinking process the moment a plausible answer is generated."*

### 4.2.4 Social Learning and Systemic Flaws

Over half of the educators (53%) saw value in ChatGPT as an observational learning tool, even for disinterested students (Q#22). Furthermore, the case of Minkowski and Einstein (Q#25) was widely interpreted as a symptom of a systemic flaw in traditional evaluation, reinforcing the need for the very transformation that educators overwhelmingly supported. For a detailed breakdown, a full set of bar diagrams visualizing the responses to all survey questions is available in the supplementary materials (Fig. SP-1).

## 5. Discussion

This study set out to investigate the philosophical and realistic perspectives of Russian university educators on the integration of ChatGPT into the learning space. The findings paint a picture not of blind technophilia nor of reactionary panic, but of a pragmatic and conditional acceptance. Educators in our sample largely view Generative AI as a catalyst for a necessary evolution in higher education—a force that exposes systemic weaknesses and demands a redefinition of pedagogical practices. The following discussion interprets these findings through the lens of our theoretical model, addresses the critical paradoxes that emerged, and proposes a concrete path forward.

## 5.1 Reimagining Assessment: Moving Beyond the "Banking Model"

The most resounding finding of this study is the overwhelming support (71%) for transforming assessment methods. The endorsement of open-book exams followed by oral dialog is a direct repudiation of what Paulo Freire termed the "banking model" of education. Our results indicate that educators recognize the futility of assessing a student's ability to act as a mere repository of information in the age where that information is instantly accessible. As one educator noted, *"The goal is no longer to see what they have memorized, but to see what they can do with the information available to everyone."*

This shift aligns with global calls for authentic assessment but is given unique impetus by Generative AI. The tool does not just make cheating easier; it makes *bad assessment* more obvious. The historical case of Minkowski and Einstein, interpreted by our respondents as a systemic failure, underscores that this is not a new problem, but rather an old one that AI has brought to a crisis point. The transformation advocated by these educators is not merely a logistical change but a philosophical one, moving the focus from knowledge retention to skills application, critical synthesis, and verbal defense—competencies that AI augments rather than replaces.

## 5.2 The Evolving Educator: From Sage to Guide and Co-Learner

A pivotal, and perhaps counter-intuitive, finding is that educators do not perceive ChatGPT as an existential threat to their profession. The negative flag for the "importance of educators" indicator reveals a resilient professional identity. Educators intuitively understand the shift predicted by our Actor-Network Theory lens: their role is being reconfigured, not diminished. The qualitative data confirms this, showing a self-perception moving from the "sage on the stage" to the "guide on the side."

This evolved role encompasses several critical functions:

- **Designer of AI-Enhanced Experiences:** Creating prompts, tasks, and projects that leverage AI for deeper learning.

- **Facilitator of Critical Dialogue:** Leading discussions where AI-generated content is critiqued, verified, and refined.

- **Mentor for Ethical and Effective Use:** Teaching the digital literacy and ethical frameworks required to use AI responsibly.

- **Assessor of Complex Competencies:** Evaluating the higher-order thinking skills demonstrated in the process of using AI tools.

As one respondent puts it, *"ChatGPT can give students the answer, but I teach them what question to ask, and what to do once they get it."* This reflects confidence that the most valuable aspects of teaching—mentorship, inspiration, and critical guidance—remain strictly in the human domain.

**5.3 The Critical Thinking Paradox: Threat or Catalyst for Deeper Learning?**

The most significant concern identified was the impact on critical thinking, the sole cognitive indicator to receive a negative flag. This presents a core paradox: while educators acknowledge AI's ability to facilitate knowledge transfer, they fear it may stifle the essential cognitive process of critique. This finding resonates with the "friend/foe" paradox in the literature.

However, this perceived threat can be reframed as a catalyst. The presence of a seemingly authoritative, yet often flawed, source of information *creates a powerful need for critical thinking*. The challenge, therefore, is not to ban the tool to protect a pristine intellectual environment, but to deliberately design pedagogy that forces engagement. This includes:

- **"Critique the AI" Assignments:** Having students analyze and improve AI-generated essays.
- **Process-Oriented Assessment:** Grading the iterative process of prompt refinement, source verification, and output synthesis.
- **Metacognitive Exercises:** Requiring students to reflect on how the AI influenced their thinking and where they had to apply their own judgment.

In this view, ChatGPT becomes a "sparring partner" for the mind, its very limitations providing the friction necessary to develop sharper critical faculties.

**5.4 Infrastructural Equity and the Risk of a New Digital Divide**

The strong conditional support for plagiarism detectors and enhanced AI features points to a pragmatic awareness of practical and ethical infrastructure required for successful integration. This extends beyond software to human capital. Our Complex Dynamic Systems framework suggests a serious risk: the **Matthew Effect**.

Students with high initial motivation, digital literacy, and access to premium AI tools will likely leverage ChatGPT to accelerate their learning, creating positive feedback loop that enhances their engagement and performance. Conversely, less prepared students may use the tool superficially, for plagiarism, or become dependent, exacerbating their disadvantages. This threatens to create new digital divide, not of access, but of *efficient use*. One educator's comment was prescient: *"We risk creating a two-tier system: one group of students using AI to fly, and another using it to crutch."* Mitigating this requires institutional investment not just in detection software, but in universal digital literacy training and support, ensuring equitable access to the skills needed to thrive in an AI-augmented learning environment.

**5.5 A Proposed Framework for Integration: The PIPE Model**

Based on our findings, we propose a practical framework to guide the responsible integration of Generative AI in higher education. The **PIPE Model** consists of four interconnected pillars:

1. **Pedagogy:** The core driver must be a shift to assessment-for-learning and pedagogy that prioritizes critical thinking, creativity, and collaboration. This is the ultimate defense against misuse and the key to unlocking AI's potential.

2. **Infrastructure:** Institutions must provide the technological and support infrastructure, including robust AI detection tools, access to advanced AI models, and AI-enabled learning platforms.

3. **Policy:** Clear, consistent, and educative policies are needed. These should move beyond simplistic bans to define responsible use, address academic integrity transparently, and outline data privacy protocols.

4. **Education (Professional Development):** Continuous, mandatory training for educators is the linchpin. This must go beyond tool usage to cover the pedagogical redesign, ethical considerations, and strategies for fostering critical AI literacy in students.

The PIPE Model emphasizes that these elements are not sequential but concurrent and interdependent. Successful integration requires pressure and progress on all four fronts simultaneously.

**5.6 Limitations and Avenues for Future Research**

This study has several limitations. The sample, while valuable, is from a specific national context, and the findings may not be fully generalizable. The "flagging" system, though effective for identifying trends, is a simplified metric. Furthermore, the study captures a snapshot in time; educator perceptions and the technology itself are evolving rapidly.

Future research should therefore:

- Conduct **longitudinal studies** to track how educator attitudes and practices change with increased exposure to AI.
- Employ the Learning Space model in **different cultural contexts** to enable comparative cross-national analysis.
- Utilize **mixed-methods approaches** with in-depth interviews to explore the nuanced reasons behind the concerns about critical thinking.

- Investigate the **student perspective** in tandem with the educator view to gain a holistic understanding of the AI-augmented classroom dynamics.

## 6. Conclusion

In conclusion, the journey forward with Generative AI in education is not about preventing a collapse, but about managing a transformation. For the educators in this study, ChatGPT is a mirror, reflecting the long-standing flaws and untapped potentials of the educational system. It amplifies the need for critical thinking, exposes the weaknesses of traditional assessment, and redefines, but does not diminish, the role of the educator. The message from the frontline is clear: the disruption is an opportunity. By embracing the challenge—through pedagogical courage, institutional support, and a renewed commitment to the human core of education—we can steer this transformation towards the future where artificial intelligence serves to augment, rather than destroy, the deeply human goals of teaching and learning.

# Supplementary Materials

This material discusses some fundamental definitions of the theoretical model of Learning Space and exemplifies with utmost details.

**Subspace:** A Subspace constitutes a major partition within the educational ecosystem, derived from an empirical definition of a domain within the "learning space" context. As multiple domains exist, the model is composed of multiple subspaces.

**Identifier:** Within the theoretical model of "Learning Space", an **Identifier** is a key constituent or a fundamental component of a specific subspace. It **identifies and embodies a set of special, defining characteristics** pertaining to a core concept, actor, or process within that part of the learning ecosystem that can be directly influenced by the integration of a new element like ChatGPT. Identifiers are the "what" that are being analyzed for impact and are measurable through the **indicators**.

**Indicator:** Indicator as a measurable variable that tracks how a new force (like ChatGPT) affects those specific characteristics.

Table S-1: Representation of the lists of the identifiers for each subspace as defined in our study.

| Subspace | Identifiers (The Key Constituents/Aspects) |
|---|---|
| **1. Learning Theory and Andragogy** | • **Social Learning Theory (SLT)**<br>• **Stage-1 (Cognitive) Learning**<br>• **Stage-2 (Associative) Learning**<br>• **Stage-3 (Autonomous) Learning** |
| **2. Complex Dynamic System of Engagement Variables** | • **Affective Engagement** (e.g., emotion, confusion)<br>• **Cognitive Engagement** (e.g., critical thinking, comprehension)<br>• **Behavioural Engagement** (e.g., student activity, click-stream data)<br>• **Academic Engagement** (e.g., grades, completion rates)<br>• **Contextual/Demographic Variables** (e.g., age, prior experience) |
| **3. Stakeholders** | • **The Learner / Educand**<br>• **The Educator**<br>• **Facilitators** (e.g., labs, libraries, MOOC platforms, search engines)<br>• **Generative AI (ChatGPT)** (positioned as a new, non-human facilitator)<br>• **Discussion Forums** (classroom and online) |
| **4. Infrastructural & Contextual Facts** | • **Digital & Physical Devices** (laptops, mobiles)<br>• **Connectivity** (internet access)<br>• **ChatGPT itself** (as a required piece of infrastructure) |
| **5. Assessment and Evaluation** | • **The Traditional Evaluation Process** (e.g., closed-book exams)<br>• **Transformed Assessment Methods** (e.g., open-book, oral exams) |
| **6. Dark Sub-Space** | • **Unethical & Prohibited Activities** (e.g., plagiarism, cheating) - collectively, this subspace *itself* is the identifier for negative outcomes. |
| **7. Null Sub-Space for Myth** | • **Prevalent Myths** (e.g., "ChatGPT will ruin students' quality") - collectively, this subspace *itself* is the identifier for debunking misconceptions. |

**Table S-2: Mapping of Identifiers to Their Measured Indicators (What It Identifies)**

| Subspace | Identifier(s) | What It Identifies / Pinpoints for Analysis |
|---|---|---|
| **Learning Theory & Andragogy** | Social Learning Theory (SLT) | The specific **psychological and pedagogical mechanism** (observation & imitation) through which students might learn from interacting with an AI model. |
| **Complex Dynamic System** | Affective, Cognitive, Behavioural, Academic Engagement | The different **dimensions of a student's relationship with learning** (their emotions, thoughts, actions, and performance) that AI could alter. |
| **Stakeholders** | The Learner, The Educator, ChatGPT (as a facilitator) | The **key actors and their interrelationships** that are being reconfigured by the introduction of a new, non-human actor into the educational network. |
| **Infrastructural & Contextual Facts** | Digital Devices, Connectivity, ChatGPT itself | The **basic hardware, software, and connectivity** that form the essential pre-conditions for using AI in education, highlighting potential equity gaps. |
| **Assessment & Evaluation** | Traditional Evaluation Process, Transformed Assessment | The **methods and philosophies for measuring learning**, pinpointing the conflict between old (rote-memorization tests) and new (application-focused) paradigms. |
| **Dark Sub-Space** | Unethical Activities | The **potential for misuse and academic dishonesty** that is inherent in the technology, identifying a risk zone that requires mitigation. |
| **Null Sub-Space for Myth** | Prevalent Myths | The **societal anxieties and misconceptions** about AI that are not grounded in reality or a nuanced understanding of education, identifying a zone of "noise" to be cleared. |

**Table S-3: Survey Questions for Educators**

| Questions Preceded by the Serial Numbers Assigned in Survey | Field Name of Dataset | Favorable Answers | Indicator Flag (50% => 0, > 50% =>1, <50% => -1) |
|---|---|---|---|
| 15) Do you vote for open book exam followed by an oral exam? | Openbook_Exam | Yes (71%) | 1 |
| 16) Do you welcome ChatGPT in education if you are facilitated with plagiarism checker that works as per your expectation? | ChatGPT_Education_Integration | Yes (61%) | 1 |
| 17) If Chatbot & the other AI tools are incorporated in educational system to build an intelligent tutoring system, would the educators lose their importance? | Educator_would_lose_importance | No (29%) | -1 |
| 18) A back-bencher can pass an exam with a probability exactly 0.5 in his last chance of retake in final semester. Will you let him to use ChatGPT in exam to get just pass mark? Consider that your institute has left it as your choice. | in_Exam_4_Non_Per | Yes (53%) | 1 |
| 19) Can use of ChatGPT harm a motivated and good student in any way? | ChatGPT_Harms_Motivated_Students | No (59%) | 1 |
| 20) During the use of ChatGPT, sequential chats may influence the prior path of knowledge when learning new things. E.g., learning quadratic function may influence some concept pertaining to linear function. Do you agree? | Backward_transfer | Yes (72%) | 1 |
| 21) There is a saying that using ChatGPT kills the critical thinking ability. In contrary, there is another saying that during sequential chats the critical thinking ability is supported and dragged ahead to be empowered. What do you say? | CHatGPT_Promotes_Critical_Thinking | Critical thinking ability empowered (41%) | -1 |
| 22) An adult student who is somehow decided not to learn a subject for some reason, ChatGPT is good for him as when he copies homework from ChatGPT, at least he would learn some terminologies which might grow his interest for that subject, otherwise he would fail and drop out at a certain point of time. | ChatGPT_Promots_SLT | Yes (53%) | 1 |
| 23) As an assistance, ChatGPT should have the facilities of diagrammatic representation and video representations. | Graphical_extension_req | Yes (73%) | 1 |
| 24) History tells us that whenever some new discoveries attempt to alter some traditional notion, a group of people raise an alarm that all is lost. But in reality, nothing is lost ever | Nothing_is_lost | Agree (76%) | 1 |
| 25) Minkowski's early disregard for his student Albert Einstein reveals the flaw persisting with the | Minkowski_Einstein | Flaws in evaluation system (28%) | 1 |

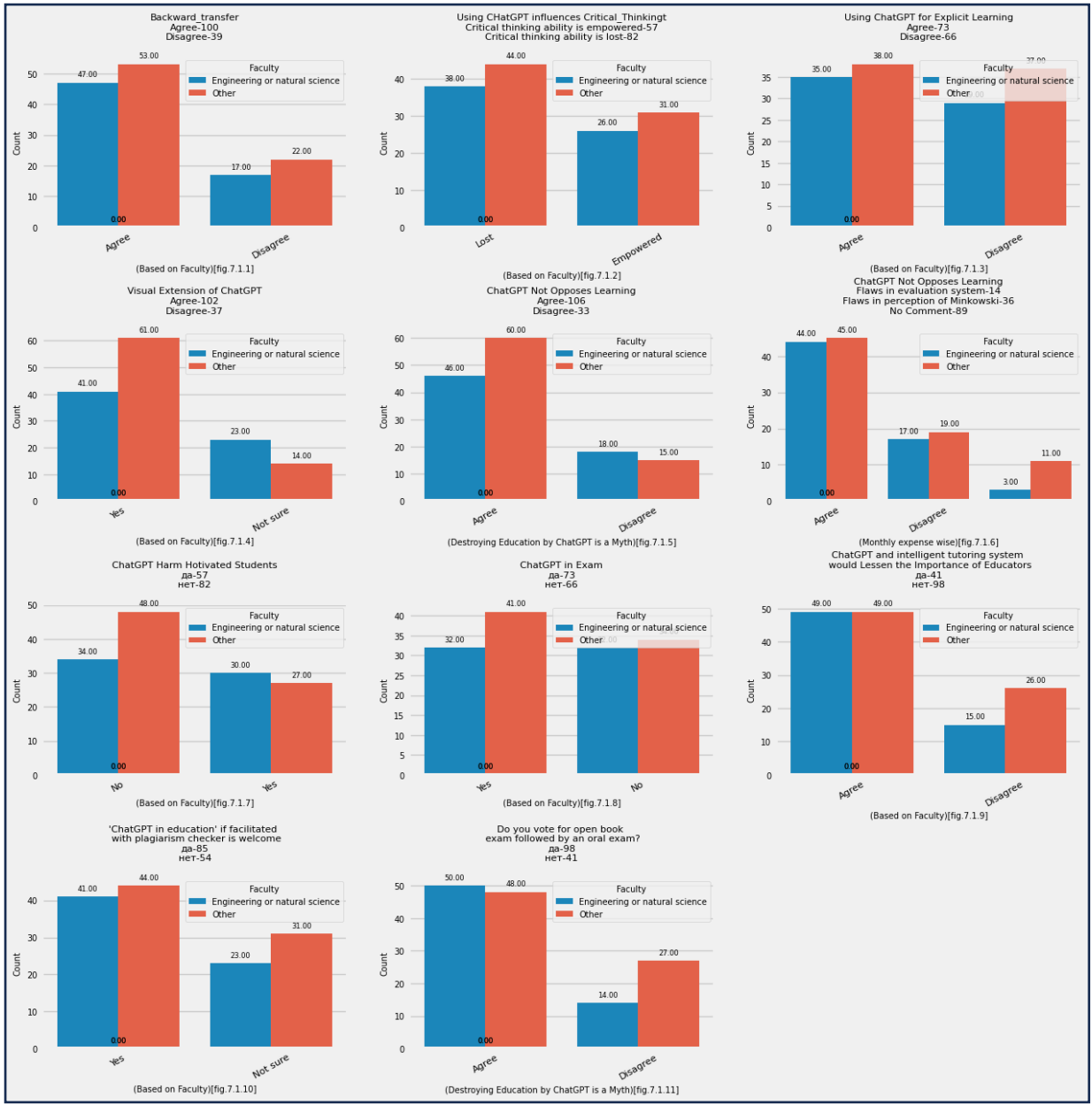

Fig. SP-1: **A grouped bar chart is provided for each of the questions, numbered from 7.1.1 to 7.1.11.**